# Exploring the Impacts of Land Use/Cover Change on Ecosystem Services in Multiple Scenarios --The Case of Sichuan-Chongqing Region, China

## 1.Introduction

Food is the most fundamental material for the survival of mankind and epochmaking for ensuring the safe and stable development of the country and society, which is vital for the all human beings (Zhang & Dong, 2023). However, to improve the quality of the environment in ecosystem, China implemented a policy of Green-for-grain project (GFGP) in 1999(T. Chen et al., 2022; Geng et al., 2019; Wang et al., 2011). The implementation of this policy has changed the way land is utilized. By 2020, a large amount of arable land in the Sichuan and Chongqing regions (SCR) and across China will have been converted to forest and construction land, leading to a significant drop in food production. D'Amour predicted that 25% of the total global arable land loss in 2030 will occur in China(Bren et al., 2017). We need to take action as soon as possible to mitigate the sharp conflict between ecology and food economy, and reduce the effect on the overall ecosystem services. However, most of the ecosystem service research in SCR focuses on ecological indicators(Cui et al., 2022; Li et al., 2023; Zhang et al., 2022; Zhong et al., 2023), and few people pay attention to the contradiction between "ecology-economy", "ecology-food". This is a research gap that cannot be ignored.

Ecosystem services (ES) are the functions of various ecosystems that are relevant for the survival and activities of mankind(Xiao et al., 2023),and ES directly and indirectly benefit human beings, forming a link between nature and human beings(Song & Deng, 2017). The current study concluded that there is a nonlinear relation between different ecosystem services(Li et al., 2022); Gu believed that soil conversation(SC) showed a decreasing trend when crop production(CP) showed an increasing trend(Gu et al., 2022); Zhang et al. concluded that water yield(WY) and CP were negatively correlated (Zhang et al., 2023); Pan et al. found that a high value of WY can promote the improvement of SC function(Pan et al., 2020). Food production and ecological protection are not a matter of extreme trade-offs, but two aspects of a complete ES system that can only be considered comprehensively by focusing on multiple indicators of ecosystem service trade-offs. As a vast grain production and reserve base in western China, The SCR has ecological characteristics such as diverse topography(Feng et al., 2023), rich vegetation types, severe soil erosion, and high ecological sensitivity(Peng et al., 2023). At the same time, its food production and arable land protection issues have a significant impact on ensuring the long-term development of national agriculture(Yin et al., 2023).Therefore, this study will provide a comprehensive assessment of CP, SC, HQ, WY and CS ecosystem services in Sichuan and Chongqing to reconcile ecological conservation with food production.

Land use and land cover change (LULC) gets a large effect on ecosystems(Qiao et al., 2021; Wang et al., 2023; Zhang et al., 2017), and it is an important way in which human activities affect ecosystems(Wang et al., 2022). Studying the effects of LULC on ES is an indispensable process for analyzing ES changes(Hasan et al., 2020; Wang et al., 2019; Zhao et al., 2022a). To better assess and simulate LULC, researchers have developed a number of studying models, incorporating the cellular automata (CA) model(Tobler, 1979), the FLUS model(Liu et al., 2017), and the PLUS model(Liang et al., 2021; Lin & Peng, 2022; Shi et al., 2021). After validation, the PLUS has been shown to be a superior model to CA and FLUS(Cao et al., 2022), which can produce more accurate simulation results and is more precise for exploring the influence of the behavioral activities of humankind on future land use(Liang et al., 2021; Wang & Zhang et al., 2022).In addition, ecological models are constantly updated and improved, among which the Integrated Valuation of Ecosystem Services and Trade-offs (InVEST) model shows better integration of ecological procedures , good spatial presentation, with high accuracy, and the model rationality is well verified(Fang et al., 2022; Li et al., 2021)

Currently, few studies have moved from a single ecological or food perspective to a coordinated "eco-food" perspective. Considering future ES projections from the perspective of balancing food production and ecological conservation is an important research gap. Therefore, we built a framework to predict future ES from the perspective of balancing food production and ecological conservation. (1). This study will use the InVEST to comprehensively evaluate the historical transformations in crop production (CP), water yield (WY), soil conversation (SC), habitat quality (HQ) and carbon storage (CS) in the study area, and to offer a preliminary understanding of some of the problems that exist in the development of the SCR. (2). On this basis, four future scenarios were constructed based on the two perspectives of "ecology-food" using the PLUS model to predict the future ES distribution. (3). We use spatial autocorrelation analysis to assess and discuss future changes in ES indicators. Through this framework, we seek the ecological-economic synergistic development planning scenario under ecological protection, which is expected to supply a scientific evidence for national land spatial programme and the delineation of the three lines, as well as to provide a leading and demonstrative role for other cities.

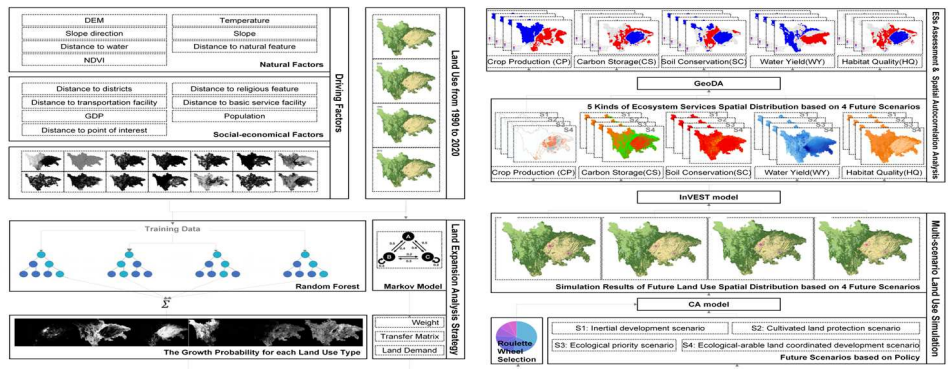

**Fig.1.** Research Framework

**2.Study Area and data sources**

*2.1 Study Area*

The Sichuan-Chongqing area includes Sichuan Province and Chongqing City. The geographical location is 97°21'~110°11'E, 26°03~34°19'N, the elevation is 77~6945 m(Xu Bin, 2022), and the total area is about 56.84×104 km². The western part of the region is an alpine plateau, and the southwestern part is mountainous with higher terrain in the west; the eastern part is a basin with lower terrain. (Fig. 1). The climate region is significantly varied. Alpine plateau area in western Sichuan features a dry and cold temperate-subarctic climate, with an basic temperature ranging from 2 to 8 ℃ in a year, average annual precipitation between 600 and 800 millimeters. The mountainous area of southwest Sichuan has an average temperature greater than 10°C in January, annual precipitation of 1000 millimeters, and a rainy season from May to October. The eastern basin has an average annual temperature of 16 to 18 degrees Celsius, an average temperature of 5~8 °C in January, an average temperature in July is 25~29 °C, and an annual precipitation of 1000 to 1200 millimeters.(Xu Bin, 2022). The soil types are complex and manifold, prevailingly purple soil, red soil, yellow soil, red-brown soil, and alpine meadow soil. (Peng et al., 2019). The plant cover and vegetation consists of subtropical broadleaved evergreen forests, subtropical coniferous forests, subalpine dark coniferous forests, montane sclerophyll evergreen living forests, subalpine shrubs, and subalpine and alpine meadows.

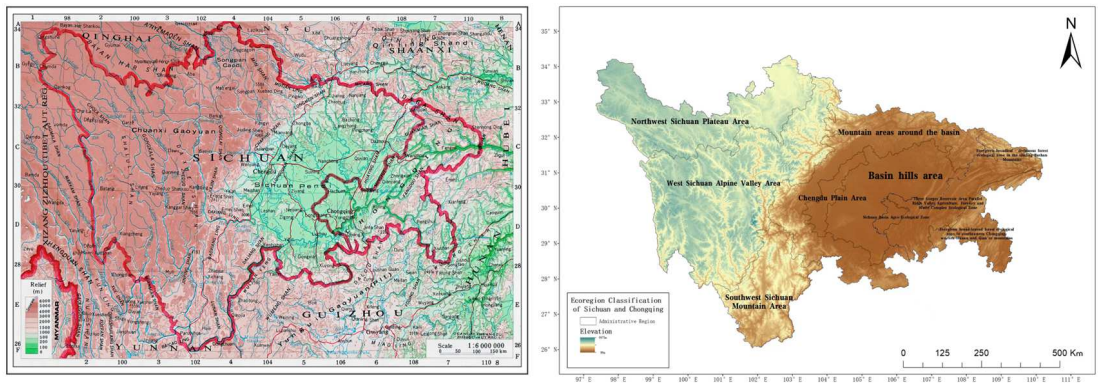

**Fig.2.** Location and topography of Sichuan-Chongqing area

2.2 Data Sources

LULC change is a complicated dynamic procedure driven by multitude of factors like natural factors, economic factors, social factors, policy factors, etc. Concerning the relevant research on land use simulation, this article chooses 14 driving factors from three aspects: natural (NDVI, Temperature, slope, etc.), economy (population, GDP), society ( distance to transportation facility, etc.), and divides the land into six types(Liu et al., 2023): arable land (AL), forestland (FL), water area (WA), grassland (GL), construction land (CL), and unused land (UL). On the basis of the PLUS model's demand for input data, all data are unified with the data as the basis for the unified coordinate system, and unified resolution of all data to 100m x 100m. and the number of rows and columns of data is strictly guaranteed.

Table 1. Details of LULC sources and factors affecting land use change.

| Data type | Data name | Data sources |
|---|---|---|
| Natural Factors | Land use and land cover | Resource and Environment Science and Data Center (RESDC)(https://www.resdc.cn/) |
| | Digital elevation model | RESDC (https://www.resdc.cn/) |
| | Slope direction | RESDC (https://www.resdc.cn/) |
| | Slope | RESDC (https://www.resdc.cn/) |
| | NDVI | MOD13Al Version 6 product (https://search.earthdata.nasa.gov/) |
| | Temperature | RESDC (https://www.resdc.cn/) |
| | Distance to water | Open streetmap (OSM)(https://www.OSM.org/) |
| | Distance to natural feature | OSM(https://www.OSM.org/) |
| Social-economical Factors | Distance to districts | OSM(https://www.OSM.org/) |
| | Distance to religious feature | OSM(https://www.OSM.org/) |
| | Distance to point of interest | OSM(https://www.OSM.org/) |
| | Distance to transportation facility | OSM(https://www.OSM.org/) |

| | |
|---|---|
| Distance to basic service facility | OSM(https://www.OSM.org/) |
| GDP | RESDC (https://www.resdc.cn/) |
| Population | RESDC (https://www.resdc.cn/) |

**3.Methodology**

Incorporating three main aspects is the proposal presented in this study for the integrated system.

1.Through the study of space and space evolution of five regions from 1990 to 2020, some problems in the regional development of Sichuan and Chongqing are preliminarily understood.

2.The PLUS model is used to simulate and forecast four scenes respectively, and the results are compared.

3.The study used the InVEST model to evaluate 006hanges in five significant ecosystems in 2050, and spatial autocorrelation analysis was used to analyze and discuss.

*3.1 INVEST ecosystem services calculation*

*3.1.1 Crop Production (CP)*

Referring to the calculation principle of the regression model in the crop production module which can be found in the InVEST model, for ten major crops whose yields were simulated globally by Mueller et al. (2011), crop production regression models can provide yield estimates for a given fertilizer input. These crops include barley, corn, wheat, etc. The model results in simulated and observed crop yields and nutritional value. All required data is based on the InVEST User Guide.

*3.1.2 Soil Conservation (SC)*

Based on the DEM-grid, the Sediment Delivery Ratio (SDR) module is utilized to calculate the erosion ratio of every grid cell by analyzing annual soil erosion of each grid cell.

InVEST's SDR module was used to build the Soil Conservation (SC) module, which is the model underlying the modeling of sedimentation. The potential for soil loss is different.

$$S_i = R_i K_i LS_i \cdot SDR_{bare\_i} \quad (1)$$

$$E_i = R_i K_i LS_i C_i P_i \cdot SDR_i \quad (2)$$

$$SR_i = S_i - E_i \quad (3)$$

According to this equation, $S_i$ is the probable sediment, $E_i$ is the real sediment, and $SR_i$ is the sequestration of a raster unit. $R_i$ refers to the rainfall scour factor of a raster; $K_i$ refers to the scour factor of a raster; $LS_i$ is the slope and slope length of a raster; $C_i$ is the crop management factor of a raster unit; $P_i$ is the corresponding practice factor of a raster; $SDR_{bare\_i}$ is the $SDR$ of a raster unit. The calculation is based on a paper by Zhao et al. (Zhao et al., 2022b).

*3.1.3 Habitat quality (HQ)*

InVEST's habitat quality model relates the source of threat in each area to the level of danger from the outside world to obtain the quality of the habitat and thus assess it (Aneseyee et al., 2020), and its habitat deterioration is carried out by the following equation.

$$D_{xj}=\sum_1^r \sum_1^y \left(\frac{\omega_r}{\sum_{r=1}^n \omega_r}\right) \times r_y \times i_{rxy} \times \beta_x \times S_{jr} \qquad (4)$$

$$i_{rxy}=\begin{cases} 1-\left(\frac{d_{xy}}{d_{rmax}}\right) \\ exp\left[-\left(\frac{2.99}{d_{rmax}}\right)d_{xy}\right] \end{cases} \qquad (5)$$

According to this equation. $D_{xj}$ is the habitat degradation of the x raster in habitat type j; r is the number of sources of danger; y is the raster in the sources of danger r; $\omega_r$ in the source of danger; $r_y$ is the stress of grid y, $\beta_x$ is the disturbance of habitat. $S_{jr}$ reflects the degree of influence of each ecological environment on various risk factors; $i_{rxy}$ is the effect of the sources of danger R in grid Y on grid X; $d_{xy}$ is the spacing between the grid Y and grid X; $d_{rmax}$ is the largest risk factor. This leads to a specific formula for habitat quality.

$$Q_{xj}=H_j\left[1-\left(\frac{D_{xj}^z}{D_{xj}^z+k^z}\right)\right] \qquad (6)$$

Under this equation, $H_j$ is the extent of habitat adaptation of species *J*; *z* is a normalization constant of general 2.5; and *k* is the degree of semi-saturation, which is usually the extent of semi-saturation of the maximum habitat.

*3.1.4 Water Yeild (WY)*

At a given period was forecasted using the yield (WY) module in the InVEST model. It is considered from the viewpoint of water balance, i.e., regional water supply reduces evaporation losses in each area (Redhead et al., 2016). The method estimates the hydropower production in question. The evaporation of water from rainfall is calculated to determine the flow rate of each pixel, and then the flow rates of each sub-watershed are combined and then averaged. Second, based on the amount of surface water in other uses, the amount of available surface water can be calculated. Third, the amount of water in the power station impoundment and the energy consumed during the life cycle of the impoundment are estimated. WY per year refers to the natural vegetation blocking effect on the ground. The equation looks like this(Zhao et al., 2022a)：

$$WR_{K,x,y}=A\times P_{x,y}\times C \times R_{K,x,y} \quad\quad\quad （7）$$

According to this equation. $WR_{K,x,y}$ are the water flows generated by the grid cells (x, y) in LULC. A is the actual zone of every grid unit. $P_{x,y}$ is the annual precipitation in this model. The experiment used annual rainfall from 1990 to 2020. C is the rainfall rate. on the surface. $R_{K,x,y}$ are the percentages of surface flow intercepted by the raster units (x, y) of LULC type K.

*3.1.5 Carbon Storage (CS)*

Carbon reservoir module uses InVEST's carbon storage cells, which project the stored carbon storage concentrations onto LULC grids, which contain forests, agricultural lands and water sources. This model combines the results of the calculations into a grid that stores the storage and values, and also includes the total quantities.

The carbon stock cell divides the resources of the entire into main categories: biogenic carbon on the surface (carbon obtained from all plant matter growing on the surface), biogenic carbon below ground (carbon present in plant roots), Carbon in the soil, and carbon dioxide from dead branches, leaves, inverted or inverted trees. This model is based on several aspects.

$$C_{total}=C_{above}+C_{below}+C_{soil}+C_{dead} \qquad (8)$$

In this formula: $C_{total}$ indicates the total carbon accumulation; $C_{above}$ the accumulation of plants in surface soil; $C_{below}$ the subsurface accumulation of plants; $C_{soil}$ the typical carbon accumulation; and $C_{dead}$ the dead organic carbon storage.

*3.2 Multi-scenario setting and PLUS land use prediction model*

*3.2.1 Multi-scenario setting*

(1) Inertial development scenario (S1): This scenario is based on social and economic drivers from 1990 to 2020, regardless of government policies, and uses the PLUS model to imitate land use change under natural development, which is the base for other scenario setting.

(2) Arable Land Conservation Program (S2): Under the current global food security situation, the conservation of AL is the key to achieving sustainable development in the world. Ensure the quality and quantity of AL in the future, this paper establishes the mobility possibility matrix of AL under this scenario based on inertia development and with reference to relevant studies, achieve the strict protection of AL.

(3) Ecological priority scenario (S3): Enhancing diversity, stability and sustainability is necessary for the harmonious development of human and nature (L. Chen et al., 2022). In this paper, the conservation of land types such as FL, GL and WA are emphasized from the current ecological requirements of China. This project is based on the traditional development model, and the high ecological interests are closely regulated.

(4) Ecology-Arable Land Coordinated Development Scenario (S4): By analyzing three different simulation scenarios, the principle of protecting farmland mainly, taking into account ecology, the principle of combining land use with land utilization is put forward to achieve the best effect of LULC.

*3.2.2 Principle of PLUS model*

PLUS is a raster-based analysis of different types for sites to more accurately reflect future site changes. The PLUS model includes the Land Expansion Analysis Strategy (LEAS) model and the CA model based on multi-type Random Seeds (CARS)(LIN Tong, 2022) . The LEAS model is able to collect and sample the site development status in two phases and the expansion effect of each region, and derive the affect of each factor on development among regions, and thus the development potential of various sites in the region. The CARS model associates random seedling creation and threshold reduction methods to determine whether the overall number of land use can satisfy the development potential conditions in terms of adaptation coefficient, neighborhood avoidance effect, and development potential on a larger scale. future needs. The available literature reveals that the PLUS model is able to commendably reflect different regional use conditions, its accuracy is more accurate than different models such as FLUS (Wang & Huang et al., 2022), and has more realistic implications to better support planning strategies.

*3.2.3 LULC simulation data*

According to the PLUS model, 14 selected driving factors were input, and the LULC in 1990 ~ 2020 was studied by LEAS model, and the land use in 30 years was obtained. On this basis, Markov chain is applied to anticipate the natural development in 2050, and finally, CARS model is used to simulate the LULC change in 2050.

(1) Natural factors, elevation and others are the significant factors impacting the change of land use; economic factors: GDP and other factors; social factors: natural land, water resources, administrative units, transportation sites, public service agencies, religious land and public land.

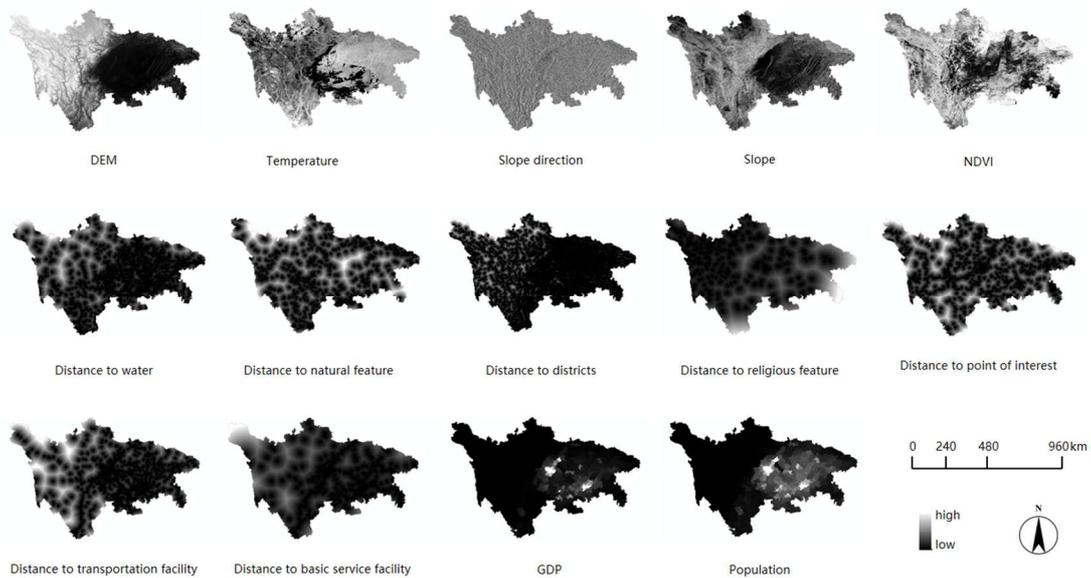

**Fig.3.** Driving factors affecting LULC

(2) Neighborhood weight setting:

The neighborhood weight was determined according to the proportion and the research experience of predecessors in this area. The final values are as follows (Table 1).

Table 2. Neighborhood factor parameters

| The type of land use | A | B | C | D | E | F |
|---|---|---|---|---|---|---|
| The neighborhood factor parameter | 0.4 | 0.6 | 0.4 | 0.4 | 0.9 | 0.8 |

A-AL; B-FL; C-GL; D-WA; E-CL; F-UL

(3) Scenario setting:

In this paper, we set up a conversion cost matrix based on four development scenarios (Table 2) for the simulation prediction of land use.

Table 3. Scenario setting

| Scenario Design | Scenario description |
|---|---|
| Natural development scenarios | The future land change pattern is consistent with the change from 1990 to 2020, and the demand for land is simulated using Markov model, and all lands are set to be interconvertible. |
| Arable land conservation scenarios | Set AL is strictly prohibited to transfer out, other land types can be converted to AL except for CL. |
| Ecological Priority Scenario | Set up FL and WA is strictly forbidden to be transferred out, GL can only be transferred out to FL and WA, and AL can be converted to any land type. |
| Ecology - Arable Land Harmonization Scenario | The land types are ranked according to their ecological benefits: FL, GL, WA, UL, AL, and CL (Li et al., 2022c); it is set that AL and Fl can only be converted to each other, and land types with lower ecological benefits than FL can be converted to AL. |

*3.3 Local space autocorrelation analysis*

Spatial autocorrelation (SA) analysis is one of the commonly used analysis methods in spatial statistics, which mainly measures the spatial correlation according to the location of features and feature values. The regional structure of each control parameter is given. At the same time, SA can also be used as an essential index to detect the relationship between a specific attribute and its adjacent spatial point. Negative relationships are just the opposite. Global autocorrelation (GA) and local autocorrelation (LA) are two directions of analysis of SA. In this method, LA is used to express the similarity between regions, to reflect the overall inclination of regions (including orientation and size), and to indicate the consistency of different spaces and the spatial dependence between different locations. The importance of the Moran I index is assessed by the values of the Moran I index, Z and p, under a range of characteristics and related attributes, usually local Moran I values. For the local spatial autocorrelation analysis, the Moran scatter plot mainly reflects the visualization degree of the spatial association mode of the indicator is divided into four types according to four quadrants, and the positive correlation type is represented by the HH type of the first quadrant (that is, the spatial unit with an attribute value above the mean is surrounded by the field with the attribute value above the mean) and the LL type of the third quadrant; The negative correlation type is represented by quadrants 2 and 4, which are represented by HL and LH types, respectively.

This paper divides Sichuan and Chongqing into 10 km × 10 km grids and analyzes the regional agglomeration of Sichuan and Chongqing by using Lisa in Li and other 2016 GIS. Its basic algorithm is like this.

$$MI_i = \frac{(X_i - \bar{x})}{S^2} \sum_i \omega_{ij}(x_j - \bar{x}) \quad (9)$$

This equation, $MI_i$ [1, 1]. Regularity represents the accumulation of similarity near a unit, while negation represents the accumulation of different spaces, indicating that there is no spatial correlation between the area and the adjacent area.

## 4. Results

*4.1 LULC and ES changes from 1990 to 2020*

*4.1.1 LULC change from 1990 to 2020*

In terms of quantitative changes, from 1990 to 2020 (Table 3), the proportion of AL decreased the most, and the proportion of FL increased the most among all types of LULC in Sichuan and Chongqing. The region of AL and GL decreased significantly, the proportion of AL decreased from 28.10% to 25.77%, and the proportion of GL decreased from 29.98% to 28.20%, decreasing by 1,783,800 ha and 1,360,500 ha respectively. The proportion of FL increased from 40.16% to 43.10%, the proportion of WA increased from 0.82% to 0.93%, the proportion of CL increased from 0.28% to 1.08%, and the proportion of UL increased from 0.66% to 0.91%, increasing by 2,254,900 ha, 843,000 ha, 614,900 ha, and 190,300 ha, respectively.

Table 4. Share and rate of change of different types of LULC in 1990 and 2020

| The type of land use | 1990 | 2020 | Change ratio |
| --- | --- | --- | --- |
| A | 28.10% | 25.77% | -2.33% |
| B | 40.16% | 43.10% | 2.94% |
| C | 29.98% | 28.20% | -1.77% |
| D | 0.82% | 0.93% | 0.11% |
| E | 0.28% | 1.08% | 0.80% |
| F | 0.66% | 0.91% | 0.25% |

A-AL; B-FL; C-GL; D-WA; E-CL; F-UL

From the spatial layout (Figure 4), by 2020, the size of site type in Sichuan and Chongqing will be FL> GL > AL > CL> WA > UL. CL is primarily expanded in the AL area, which is clustered and spread over the Chengdu-Chongqing urban agglomeration within the Sichuan basin, mainly in 16 cities such as Chongqing and Chengdu; the FL is distributed around the Sichuan basin, mainly in Liangshan Yi

Autonomous Prefecture and Ya'an City; the GL is chiefly distributed plateau area of high mountains in the northwestern Sichuan-Chongqing region, specifically in Ganzi region. The primary reasons for rapid urban development and population inflow are the vast urban land expansion in Chengdu and Chongqing throughout the previous three decades, leading to the depletion of vast amounts of AL.  Additionally, the initiative of "returning farmland to the forest" has accelerated the reduce in AL and the increase in FL.  This has resulted in the embezzlement of a good deal of AL in these cities.

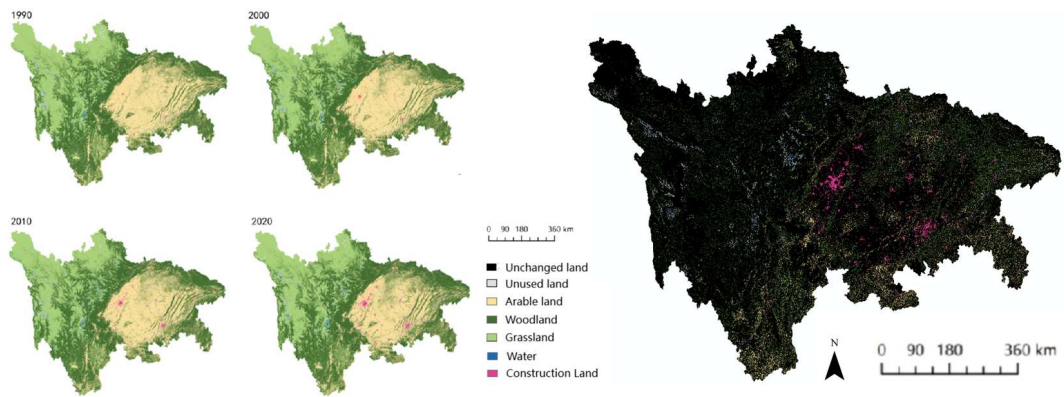

**Fig.4.** LULC change from 1990-2020 (the left graph shows the distribution of land types for four years; the right graph shows the expansion distribution of each land type from 1990 to 2020)

*4.1.2 Dimensional and time variation in ESs from 1990 to 2020*

Regarding quantity change (Figure 5), CS, SC, and HQ were constantly on the rise, with an equilibrium growth rate of 1.68%,0.08%, and 0.46%, respectively. CP gradually decreased, with an average decreasing rate of 2.75%. WY decreased first and then increased, with a decrease rate of 0.11% averagely.

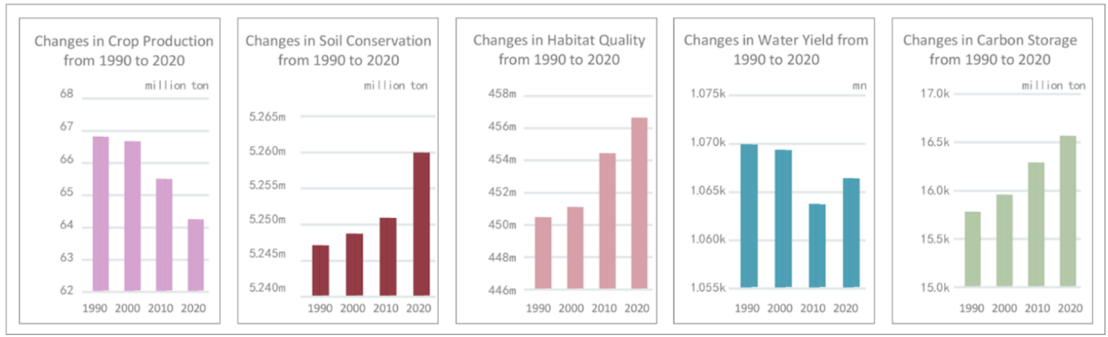

**Fig.5.** A histogram of changes in the number of ESs between 1990 and 2020

From the spatial layout (Figure 6), the high-value CS, SC, and HQ areas are concentrated in the Sichuan basin's FL areas. The FL areas have the most significant contribution to ecology, mainly in Aba, Ya'an City, Liangshan Yi Autonomous Prefecture, and some areas in Guangyuan City and Chongqing City, and the expansion trend of CS, SC, and HQ in these areas gradually increases due to the expansion of FL; the low-value regions are principal located in the Sichuan basin. And the median value of CS and HQ is located in the GL area, mainly in Ganzi and Aba, and the change over time is small, while the SC in most areas is at a low-value. The high-value areas of WY is situated in the eastern part of the region, mainly in the CL and AL in Meishan, Leshan, and the eastern area of Chongqing; the middle value is principally in the AL in the urban agglomeration of Chengdu-Chongqing and the northern area of Aba; the low-value is principally in the FL and GL areas in the high mountain plateau in the west; the distinction in WY is related to the land type and the topography of the Sichuan and Chongqing regions.CL and FL expansion in Chengdu, Meishan and Neijiang has led to a decrease in plowland in these areas. The trend of CP shrinkage is more pronounced in these areas. Low-value areas are principally located in the forest-field interval in the southeast of the AL area.

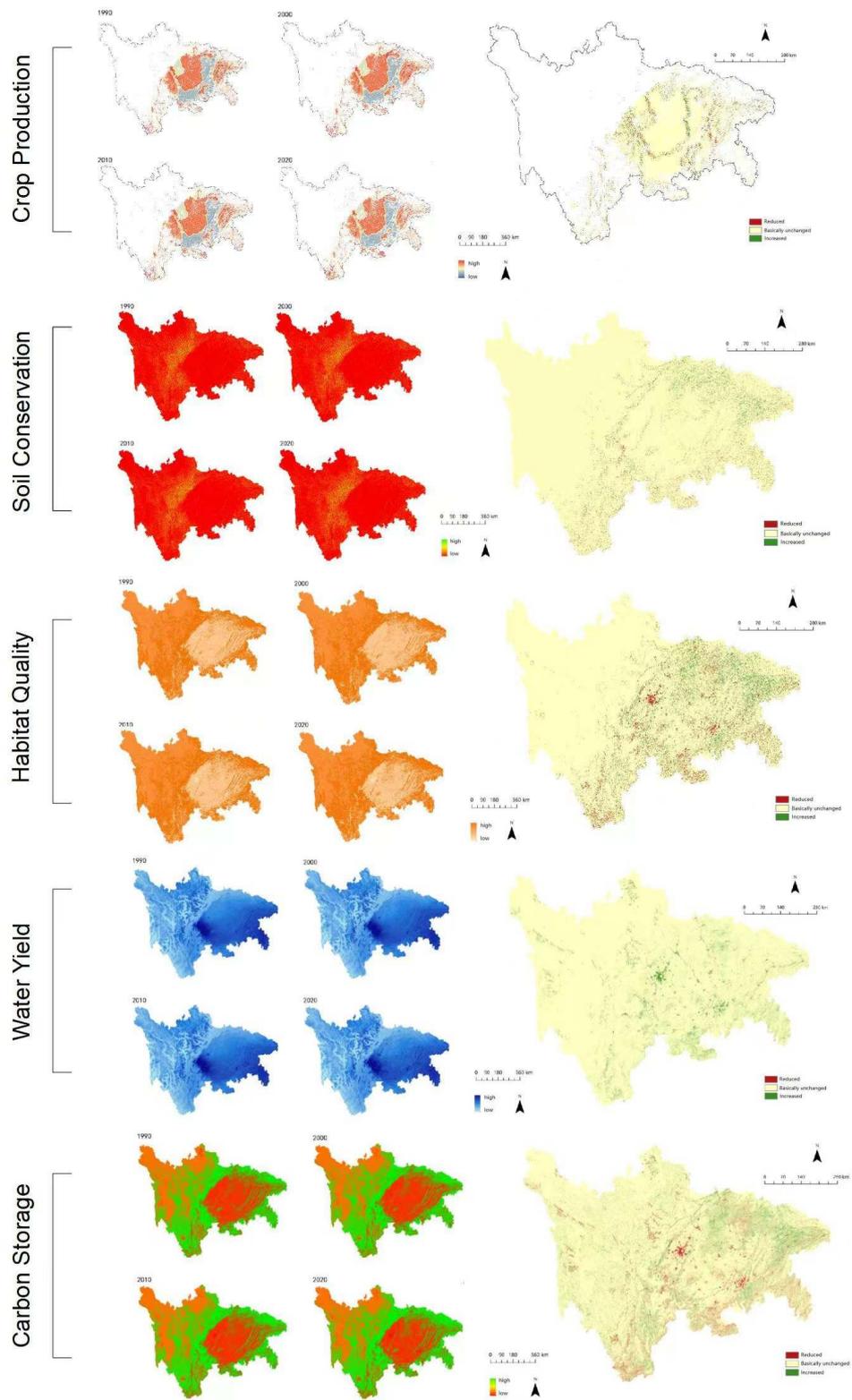

**Fig. 6.** Spatial and temporal changes in ESs from 1990-2020 (left panel shows the spatial distribution of ESs for four years; right panel shows the spatial distribution of changes in ESs from 1990-2020)

*4.2 Impacts of LULC change on ESs under different future scenarios in 2050*

*4.2.1 LULC change under different future scenarios in 2050*

In terms of quantitative changes (Table 4), CL was enlarged in all three scenarios, except for the S4 scenario, with 0.57%, 0.21%, and 0.37% expansion, respectively; the maximum ratio of CL is 1.65% in the S1 scenario, which is due to the development of urbanization in the realistic scenario. In scenarios S2 and S4, there is a slight expansion of the AL area, 0.74% and 0.08%, respectively; in scenarios S1 and S3, the decreased AL area is more significant, 1.72% and 2.93%, respectively. Since S2 is the scenario of arable land protection, the acreage of the AL area is the biggest, 26.52%. Under all four scenarios, the FL area expanded by 2.71%, 0.29%, 4.22%, and 1.41%, respectively; Under the S3 scenario, which aims at ecological conservation, the maximum acreage of FL area is 47.32%. Under all four scenarios, the GL area was reduced by 1.70%, 1.29%, 1.92%, and 1.59%, respectively. Out of the S1 scenario, the WA increased in the other three scenarios by 0.08%, 0.30%, and 0.16%, respectively; under the S3, the most significant proportion of WA was 1.23%.

Table 5. LULC volume changes in 2050 for the four scenarios (Rate of change for 2050 versus 2020)

| Scenario | S1 | | | S2 | | | S3 | | | S4 | | |
|---|---|---|---|---|---|---|---|---|---|---|---|---|
| | Land area | Proportion | Change rate | Land area | Proportion | Change rate | Land area | Proportion | Change rate | Land area | Proportion | Change rate |
| A | 18439673 | 24.05% | -1.72% | 20331563 | 26.52% | 0.74% | 17512408 | 22.84% | -2.93% | 19822000 | 25.85% | 0.08% |
| B | 35128200 | 45.81% | 2.71% | 33272795 | 43.39% | 0.29% | 36280739 | 47.32% | 4.22% | 34132000 | 44.51% | 1.41% |
| C | 20323993 | 26.51% | -1.70% | 20639191 | 26.92% | -1.29% | 20153993 | 26.28% | -1.92% | 20405490 | 26.61% | -1.59% |
| D | 689675 | 0.90% | -0.03% | 781000 | 1.02% | 0.08% | 945083 | 1.23% | 0.30% | 841732 | 1.10% | 0.16% |
| E | 1266092 | 1.65% | 0.57% | 993183 | 1.30% | 0.21% | 1116096 | 1.46% | 0.37% | 831016 | 1.08% | 0.00% |
| F | 830099 | 1.08% | 0.17% | 660000 | 0.86% | -0.05% | 669413 | 0.87% | -0.04% | 645494 | 0.84% | -0.07% |

A-AL; B-FL; C-GL; D-WA; E-CL; F-UL

Regarding spatial layout (Figure 7), under scenarios S1 and S3, CL encroaches heavily on AL within the Chengdu-Chongqing urban agglomeration in the Sichuan Basin, achieving a large area of expansion; under scenarios S2 and S4, CL extension is not apparent. AL expands significantly in S2 and S4 scenarios in the Sichuan Basin, mainly on unused and ecological land. S2 also includes expansion in FL surrounding the basin. S4 focuses on protecting original high-quality contiguous land use. In scenarios S1, S3 and S4, there is a significant expansion of FL around the Sichuan Basin, mainly encroaching on UL and GL, which follows historical land use patterns and ecological priority scenario. The WA of the Yalong and Dadu Rivers in the western alpine plateau grass area increases significantly in scenarios S2 and S4. The UL around the WA also increases significantly, and these unused lands may be swamps around the rivers, etc. In scenario S3, because the AL area can be converted to ecological land, the Yangtze, Min, and Pei Rivers in the AL area of the Sichuan Basin expand significant

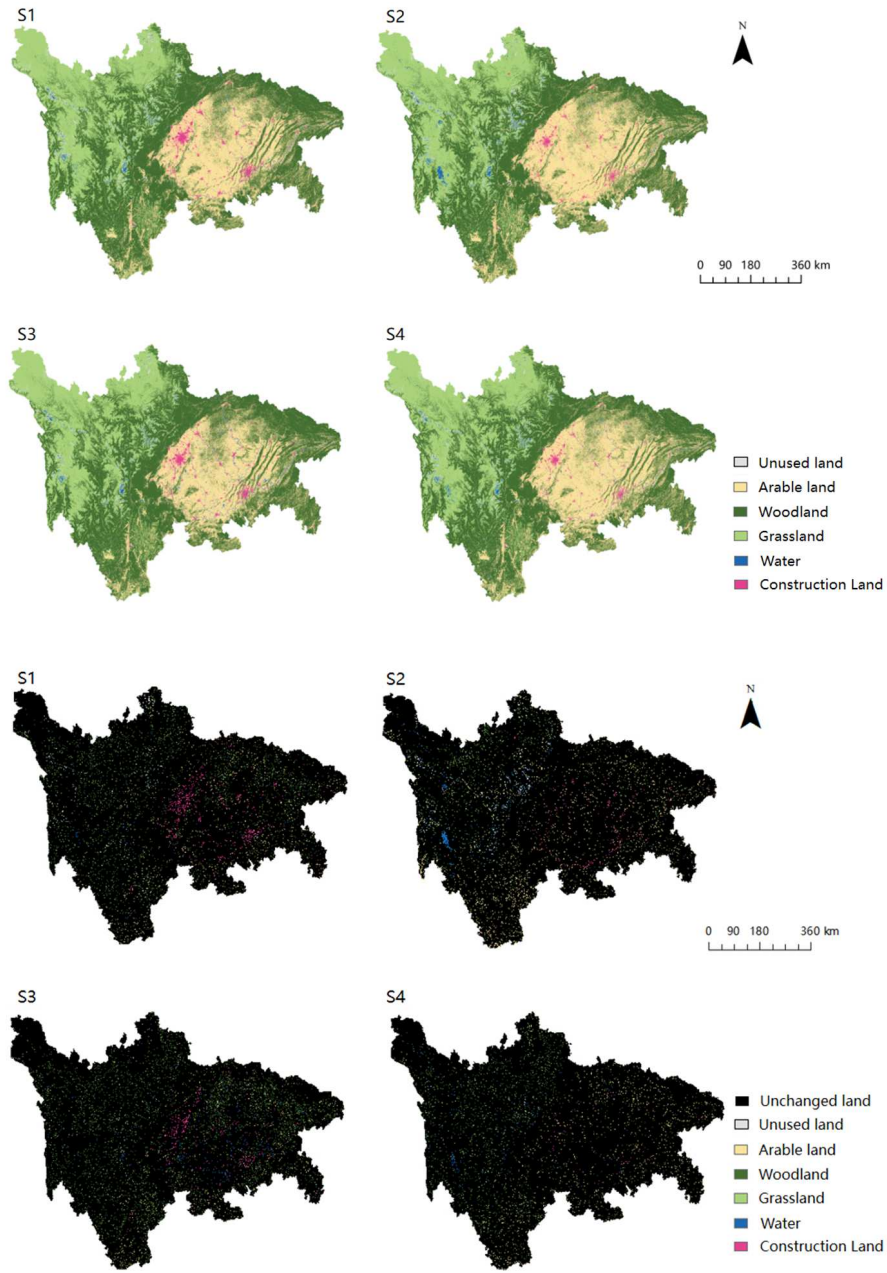

**Fig.7.** LULC change in 2050 under four scenarios (the left figure presents the land types distribution in the four cases; the right figure presents the distribution of expansion of each land type under each scenario with respect to 2020)

*4.2.2 ESs changes under different future scenarios in 2050*

The spatial distribution of ecosystem services in 2050 under the four scenarios is shown in Figure 8, Figure A, and the change curve of ecosystem services in 2050 under the four scenarios is shown in Figure B. The change curve of ecosystem services in 2050 under the four scenarios is shown in Table 5. From Table 5, the numerical changes in the conversion rate of ecosystem services from 2020 to 2050 can be obtained. The predicted results of virous scenarios show that the trend of all ecosystem service indicators in the S4 scenario is between S2 and S3. 4.22% expansion of FL in S3, and the expansion of LULC for ecology increases SC, HQ, and CS by 0.38%, 3.41%, and 7.44%, respectively. The expansion of forestlands and grasslands will increase plant transpiration to some extent, which will be shown in changes in ecosystem services as a reduction in WY. Scenarios S2 and S4 focus on conserving AL. These two scenarios result in an increase in AL area and an increase CP by 2.25% and 0.35% respectively, compared to S1. This reverses the trend of decreasing food production under S1. This highlights the significance of AL conservation in addressing the global food crisis. However, focusing on AL conservation will make in S2 scenario SC and WY decrease by 0.17% and 0.21%, respectively. Comprehensive analysis shows that the change of indicators under the S4 scenario is more balanced and meets the requirements of long-term development in the setting of the global food crisis.

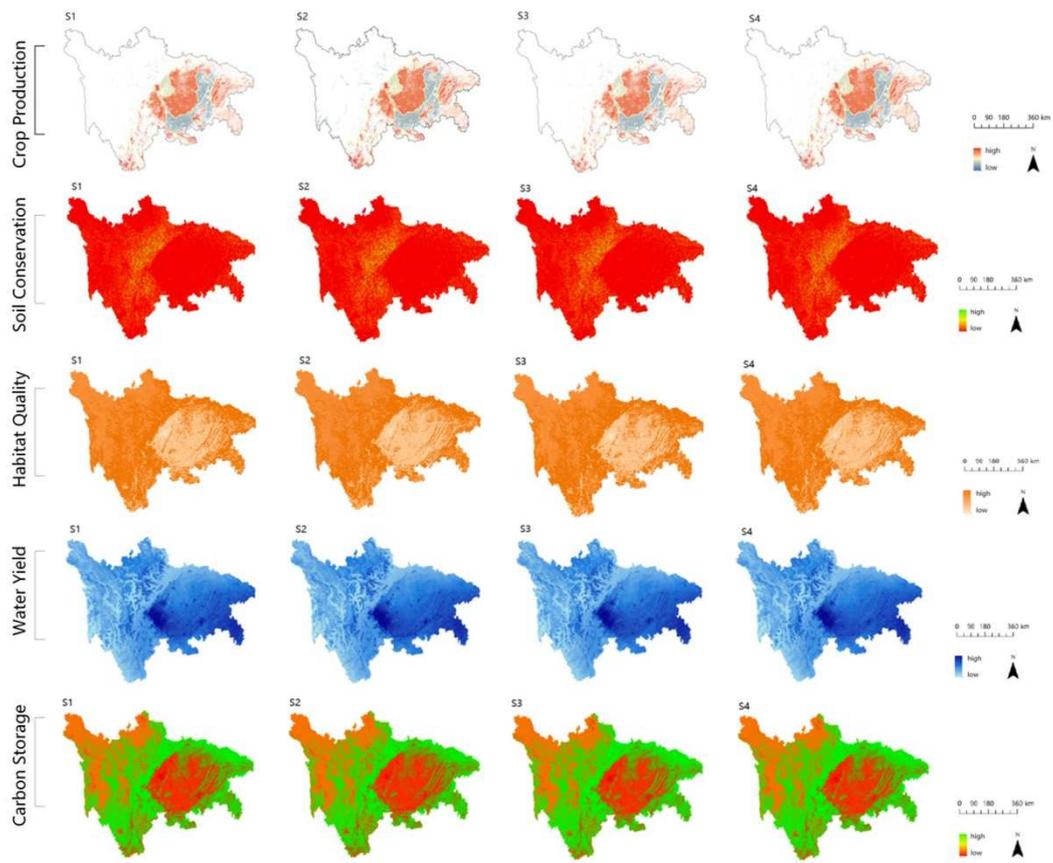

(a)

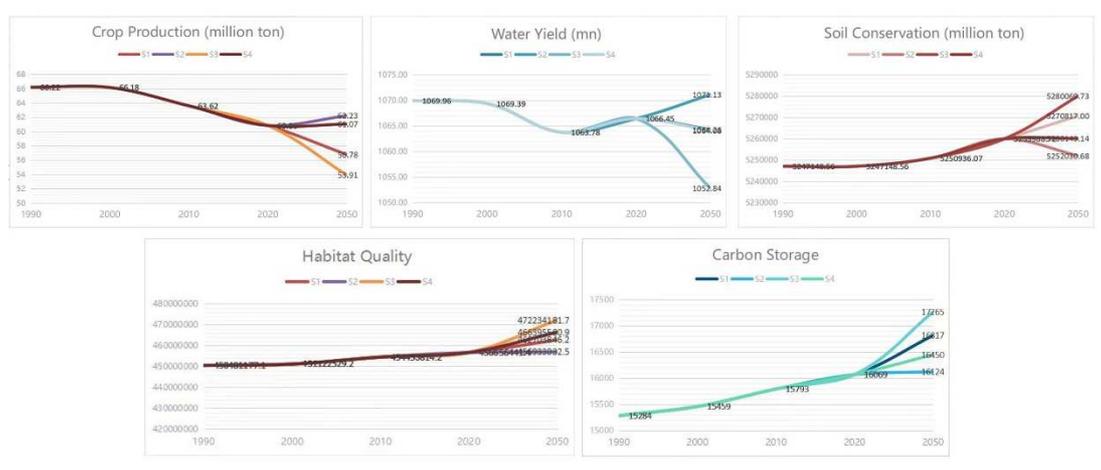

(b)

**Fig.8**. Spatial distribution of ecosystem services in 2050 under four scenarios (a) and ecosystem service change curves (b)

*Table 6* Conversion rate of ecosystem services 2020-2050

|    |    | 2020 | 2050 | conversion rate |
|----|----|------|------|-----------------|
|    | CP | 60.86 | 50.78 | -16.56% |
|    | WY | 1066.45 | 1071.13 | 0.44% |
| S1 | SC | 5259988.72 | 5270817.00 | 0.21% |
|    | HQ | 456656441.4 | 462703846.2 | 1.32% |
|    | CS | 16069 | 16817 | 4.65% |
|    | CP | 60.86 | 62.23 | 2.25% |
|    | WY | 1066.45 | 1064.26 | -0.21% |
| S2 | SC | 5259988.72 | 5251030.68 | -0.17% |
|    | HQ | 456656441.4 | 456988032.5 | 0.07% |
|    | CS | 16069 | 16124 | 0.34% |
|    | CP | 60.86 | 53.91 | -11.42% |
|    | WY | 1066.45 | 1052.84 | -1.28% |
| S3 | SC | 5259988.72 | 5280069.73 | 0.38% |
|   | HQ | 456656441.4 | 472234181.7 | 3.41% |
|   | CS | 16069 | 17265 | 7.44% |
|   | CP | 60.86 | 61.07 | 0.35% |
|   | WY | 1066.45 | 1064.08 | -0.22% |
| S4 | SC | 5259988.72 | 5260143.14 | 0.00% |
|   | HQ | 456656441.4 | 466395590.9 | 2.13% |
|   | CS | 16069 | 16450 | 2.37% |

*4.3 Spatial autocorrelation analysis of ESs in 2050*

    The Moran's I index reflected the spatial distribution of high or low ES functions in the Sichuan-Chongqing region. Fig.9 shows a local auto-correlation of the future ESs.

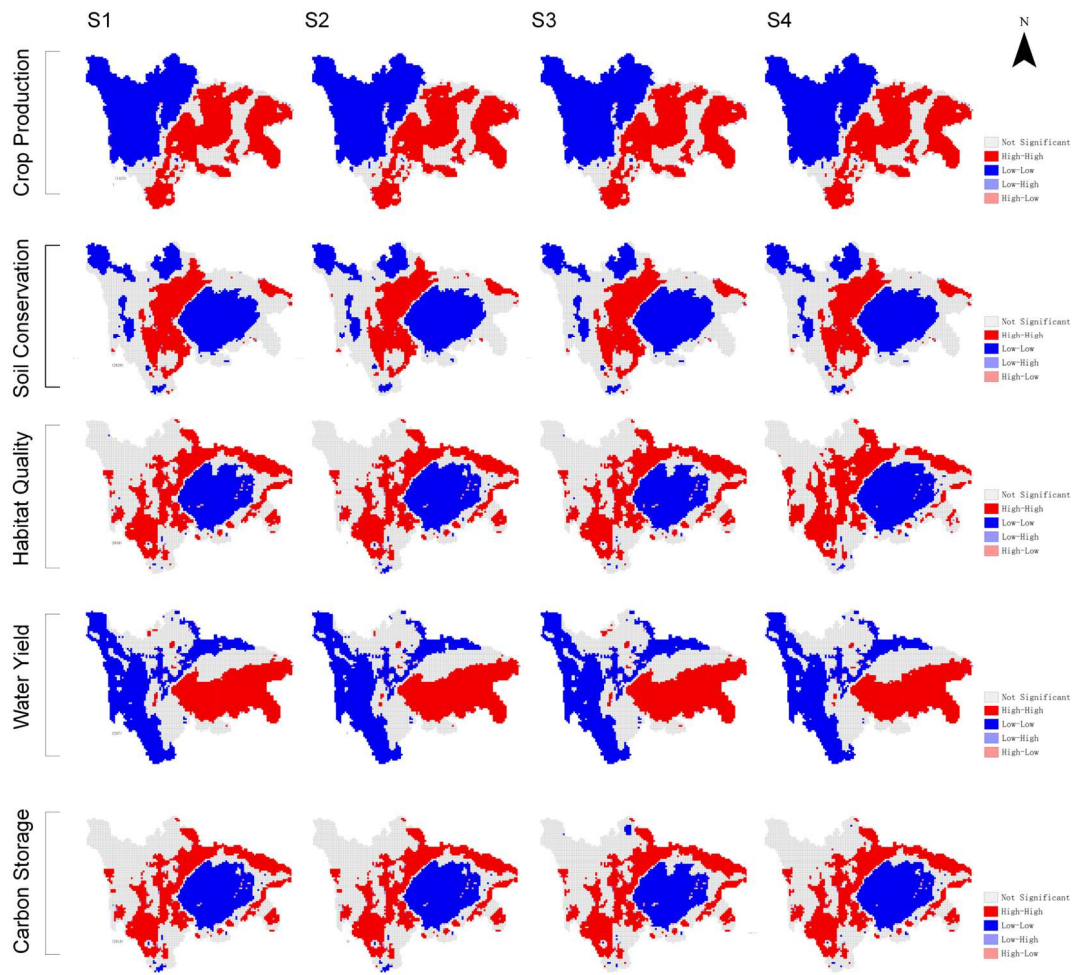

**Fig. 9.** Local spatial auto-correlation analysis of ecosystem services under

four scenarios in 2050

The relative distribution of the areas with high-high autocorrelations（HH） and the areas with low-low autocorrelations (LL) in each scene shows a certain regularity. In the area of concentrated distribution of low-altitude administrative regions in the southeast, SC, HQ, and CS all show concentrated distribution of LL, and CP shows a concentrated distribution of HH. In the sparse distribution area of high-altitude administrative regions in the northwest, LL is concentrated under CP and WY, and HH is concentrated under HQ and CS. Meanwhile, HQ and SC show high similarity in HH, LL, the areas with low-low autocorrelations (HL), and the areas with low-high autocorrelations (LH), and SC and WY show some complementary distributions in the northwest region.

For CP, HH and LL mainly thrives in northern and southern Sichuan-Chongqing areas. In the southern region, LL has the lowest value under the S3 scenario; in the northern region, HH has the highest value under the S4 scenario. Regarding SC, HH, LL, and in the S2 scenario, LH anomalies can be found dispersed throughout both central and marginal regions. The LL and HH anomalies in HQ are concentrated in the central part. In this region, the highest value of the S4 distribution and the lowest value of the S3 distribution was found for the LL distribution. For the HH distribution, the S4 distribution is the lowest, and the S3 distribution is the highest. Regarding FL, the HH anomalies are scattered, and they all show a trend in which the HH values of S4 are between those of S2 and S3. Regarding CS, the pattern exhibited between scenarios in the southeast direction was consistent with the HQ. At the same time, LL anomalies were observed on the northern side, where the LL values of S3 were higher than those of S4.

## 5. Discussion

*5.1 Impact of LULC conversion of Ess*

Based on Figure 4, the land use types in SCR have more significantly conversed during the last three decades, which is caused by the combined effect of many factors, such as urban development, changes in environmental conditions, and policy adjustments.

The impact of land use changes on the ecosystem performs much better than that of other natural elements (Li et al., 2003). The low-cost areas in CS, HQ, and SC are all located in the arable land area of the Sichuan Basin. the same conclusions have been drawn by Hu(A. Hu et al., 2022) and Bai(Bai et al., 2018). On this distribution of ecosystem services in this region. This also indicates that forestlands drive more ecosystem services in Sichuan and Chongqing, like SC, HQ and CS, which is in harmony with the conclusions obtained from the research conducted in Sichuan Province. The high-cost area in CP is situated in the cultivated area of the Sichuan Basin. Moreover, the high-cost region governed by cultivated and construction land of WY is mainly positioned in the Sichuan Basin area.

In contrast, the low-cost region is mainly positioned in the FL and GL territories in the western alpine plateau region. This is in harmony with the research results of Bai et al. (Bai et al., 2017) on water containment in Sichuan. Additionally, the annual water supply of CL and AL is greater than that of FL and GL, which is the same finding as Hu et al. (W. Hu et al., 2022).

5.2. Scenario-driven Ess

In different development contexts and policy orientations, ecosystem services are bound to change in time and space, with Sichuan Province taking the lead in "returning farmland to the forest" as a pilot project in 1999, mainly leading to a decrease in AL and CP from 2000 onwards. Green development focuses on energy conservation, emission reduction, reforestation and combating desertification. These measures have led to increased carbon storage and improved soil and habitat quality. This contributes to ecological restoration and sustainable development. However, in the absence of other policy influences, the extreme ecological priority can make the expansion of representative ecosystems like forests, grasslands, and water encroach on arable land and thus lead to lower food production. Therefore, to cope with the global food crisis, we should actively protect arable land and keep the bottom line of "returning farmland to the forest" to ensure sustainable development. In the S4 scenario set up in this paper, a certain balance between arable land conservation and ecological development is achieved, and the balanced development of all ecosystem services is better than the other three scenarios.

*5.3 Limitations and future prospects*

However, this study has some shortcomings, and future studies need further consider the mechanisms of action and adaptation strategies of climate change., population growth and other factors on the impacts of LULC and its ESs, In addition, future studies should also examine the synergistic or trade-off relationships between different ecosystem services, as well as the mechanisms of action and adaptation strategies of climate change, in order to enhance the accuracy and applicability of the model.

## 6. Conclusion

Our study builds a framework to assess the historical changes of ESs in the SCR from the perspective of balancing food production and ecological conservation, and to anticipate the SCR future ESs. The subsequent conclusions are derived:

(1) The analysis of historical LULC showed that during the 30-year period from 1990 to 2020, the proportion of AL in each type of LULC in the SCR decreased the most, from 28.10% to 25.77%. The proportion of FL increased the most, from 40.16% to 43.10%. By analyzing the historical data of ecosystem services, CS, SC, and HQ all showed an increasing trend, with an average growth rate of 1.68%, 0.08%, and 0.46%, respectively. WY showed a decreasing and then increasing trend, with an average decrease of 0.11%. CP decreased more significantly with an average decrease of 2.75%. In the case of a large expansion of ecological land, the decline in food production is much larger than the rise in ecological indicators.

(2) By calculating the ESs under different future scenarios in 2050, under S1 scenario, CP shows a decreasing trend, SC, HQ, CS and WY show an increasing trend, and CP is negatively correlated with other ecological indicators. In the S4 scenario, the decreasing trend of CP is alleviated and even slightly increased, and the increasing trend of SC, HQ, and other ES is still maintained, and the negative correlation between CP and other ecological indicators is alleviated.

(3) The projections under different scenarios showed that the ecosystem services developed in a balanced manner in the S4 scenario, which was better than the other three scenarios. In the three scenarios in 2050, CP production in S4 is 4.29 million tons higher compared to S1; and WY, SC, CS and HQ perform better in S4 compared to S2. Therefore, we argue that policy leadership that considers control can guarantee food and ecological harmonization

In summary, this experiment has achieved some results. With the strengthening of the cognizance of ecological protection and the accelerated progress of urbanization, the contradiction between arable land and ecological land use will remain prominent in the future. Therefore, this research framework helps us to seek the ecological and economic synergistic development planning program under ecological protection, and provide more scientific suggestions for the implementation of policies such as land spatial planning and three-line delineation.

Table 7. List of abbreviations

| Original text | Abridge |
| --- | --- |
| Green-for Grain Program | GFGP |
| Land use/cover change | LULC |
| Sichuan-Chongqing Region | SCR |
| Ecosystem services | ESs |
| crop production | CP |
| soil conservation | SC |
| habitat quality | HQ |
| water yield | WY |
| carbon storage | CS |
| high-high | HH |
| Low-low | LL |
| Chengdu-Chongqing economic circle | CCEC |
| woodland | WL |
| grassland | GL |
| arable land | AL |
| construction land | CL |
| water area | WA |
| unused land | UL |
| Spatial autocorrelation | SA |

of the Chengdu–Chongqing City Group *International Journal of Environmental Research and Public Health* (20, pp.).